\documentclass[12pt]{article}
\usepackage{hyperref}
\usepackage{graphicx}
\usepackage{amsmath}
\usepackage{amssymb}
\usepackage[numbers,compress]{natbib}
\usepackage{amsmath}

\begin{document}
\title{Light from Reissner-Nordstrom-de Sitter black holes}
\author{Ion I. Cot\u aescu\\ {\small \it  West 
                 University of Timi\c soara,}\\
   {\small \it V. P\^ arvan Ave. 4, RO-1900 Timi\c soara, Romania}}

\maketitle

\begin{abstract}
We derive for the first time the form of the spiral null geodesics around the photon sphere of  the Reissner-Nordstrom black hole in the de Sitter expanding universe. Moreover, we obtain  the principal parameter we need for deriving, according to our method  [I. I. Cot\u aescu. {Eur. Phys. J. C.}  (2021) 81:32],  the  black hole shadow and the related redshift as measured by a remote observer situated in the asymptotic zone. We obtain thus a criterion of detecting  charged black holes without peculiar velocities when one knows the mass, redshift and the black hole shadow.   

\end{abstract}


 \newpage
\section{Introduction}

The light emitted by black holes provides us with information about two important astronomical observables, namely the black hole shadow and redshift. The black shadow is related to the photon sphere and associate spiral geodesics produced by the black hole gravity.  For this reason  a  geometric method was widely used  for studying  the  Schwarzschild black holes in the Minkowski spacetime  \cite{Sy} or  in expanding universes \cite{SS1,SS2,SS01,SS02,SS03,SS3} as well as  rotating black holes with Kerr \cite{K1,K2} or Kerr-de Sitter \cite{KdS1,KdS2,KdS3,KdS4} metrics.   A special attention was paid to the new object M87 \cite{V1,V2} which was discovered recently \cite{A1,A2}.

However, this method is not suitable for studying the redshift which depends on the energies of the emitted and observed photons whose ratio gives information about the cosmic expansion and the possible peculiar velocity of the observed black hole.  For separating these two contributions one combined so far the Lema\^ itre rule \cite{L1,L2} of Hubble's law \cite{Hubb}, governing the cosmological effect, \cite{SW,H0,H} with the usual theory of the Doppler effect of special relativity \cite{LL} even though there are evidences that our universe is expanding. Recently we proposed an improvement of this approach replacing the special relativity with our de Sitter relativity \cite{CdSR1,CdSR2} which allowed us to find a new method of deriving simultaneously the black hole shadow and redshit as observed from the asymptotic zone of the de Sitter expanding universe \cite{CDop,CBH}. By using a suitable  code on computer \cite{Comp} we may apply this method even to the black holes moving freely with peculiar velocities. 

Our method is based on the conserved quantities on the spiral null geodesics that can be measured by a remote observer when we know the relative motion of the black hole with respect to this observer. The spiral geodesics are rolled out on the photon sphere escaping outside near their specific singularities such that the photons on this geodesics give the light around the black hole shadow defining thus its radius. A remote observer, which neglects the black hole gravity, measures an apparent genuine de Sitter null geodesic of a photon of given energy and momentum, emitted by an apparent source near the black hole shadow. Thus we may derive simultaneously the redshift related to the measured photon energy and the shadow angular radius given by the observed photon momentum \cite{CBH}.  In the case of the Schwarzschild-de Sitter black hole we met spiral geodesics which are very similar to Darwin's ones \cite{D1,D2}, derived for this black hole in the flat spacetime \cite{CBH}.  However, if we intend to extend this study to the  Reissner-Nordstrom black hole we find that only its photon sphere was studied so far \cite{RN}  but without deriving the form of the spiral geodesics and their conserved quantities. 

For this reason we would like to continue here this study focusing on  the light around the Reissner-Nordstrom black holes for deriving the  spiral geodesics and the conserved quantities we need for writing down the closed formula of the radius of the sphere hosting the apparent sources around the black hole shadow and  the observed energy giving the redshift. For this purpose we consider the Reissner-Nordstrom-de Sitter black holes in comoving local charts (frames) with  Painlev\' e coordinates \cite{Pan} that appear as genuine de Sitter {comoving} frames \cite{BD} for the remote observers.  In this lay out we derive  for the first time the equation of the spiral geodesics of the Reissner-Nordstrom-de Sitter black holes pointing out their properties and deriving the radius of the sphere of the apparent sources that may be substituted in our general results  \cite{CBH,Comp} for finding  the  shadow and redshift of the  Reissner-Nordstrom black holes moving feely in the de Sitter expanding universe. 

We start in the second section reviewing the proper comoving frames with Painlev\' e coordinates \cite{Pan} of the Reissner-Nordstrom black hole in the de Sitter expanding universe observing that for a remote observer these appear as de Sitter comoving frames. The next section is devoted to the spiral geodesics and their conserved quantities solving analytically the  equation of the null geodesics and deriving the radius of the apparent sources near the black hole shadow. In the third section we show how the redshift and black hole shadow can be studied with our general method \cite{CBH,Comp} by using the parameters derived here. Moreover, a rapid method of detecting charged black holes without peculiar velocities is proposed.  Finally, we present our concluding remarks.

In what follows we use natural Planck units with $c=\hbar=G=1$ and the notations of Ref. \cite{CBH} where $\omega_H=\sqrt{\frac{\Lambda}{3}}\,c$ is the de Sitter Hubble constant (frequency) while  the Hubble time  $t_H=\frac{1}{\omega_H}$ and the Hubble length   $l_H=\frac{c}{\omega_H}$  have the same form. 

\section{Preliminaries}

Let us start with  $(1+3)$-dimensional isotropic pseudo-Riemannian manifolds, $(M,g)$, where we may introduce local frames $\{x\}$ of coordinates $x^{\mu}$ ($\alpha,\mu,\nu,...=0,1,2,3$). For the systems with spherical symmetry we may chose spherical space coordinates,  $(r,\theta,\phi)$,  and different time coordinates. The traditional {\em static} frames, $\{t_s,r,\theta,\phi\}$, depend on  the static time  $t_s$ having  line elements of the form
\begin{equation}
ds^2=g_{\mu\nu}dx^{\mu}dx^{\nu}=f(r)\, dt_s^2-\frac{dr^2}{f(r)}-r^2 d\Omega^2\,,\label{s1s}
\end{equation}
where $d\Omega^2=d\theta^2+\sin^2\theta\, d\phi^2$. These line elements can be put at any time in Painlev\' e's forms \cite{Pan}, 
\begin{equation}
ds^2=f(r)dt^2+2\sqrt{1-f(r)}\,dtdr-dr^2-r^2 d\Omega^2\,,\label{ss}
\end{equation}
substituting in Eq. (\ref{s1s}) 
\begin{equation}\label{subs1}
t_s=t+\int dr \frac{\sqrt{1-f(r)}}{f(r)}\,,
\end{equation}
where $t$ represents the {\em cosmic} time of the frames $\{t, r,\theta,\phi\}$ which have flat space sections.   

{ \begin{figure}
 \centering
   \includegraphics[scale=0.85]{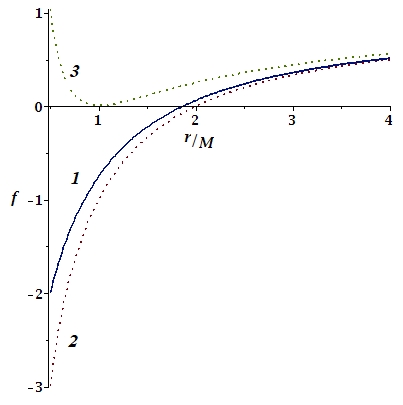}
 \caption{Function $f(r)$  near the exterior horizon  $r_+=1.8867 M$  of a black hole with  $\mu=0.01$ and $q=0.5$ (1) between the Schwarzschild-de Sitter  one  with $q=0$  (2) and that of the extremal black hole obtained here for $q=1.005$ (3). }
  \end{figure}}

The static frame of  a Reissner-Nordstrom black hole of mass $M$ and charge $Q$  in de Sitter expanding universe,  has the metric (\ref{s1s}) with 
\begin{equation}\label{fr}
f(r)=1-\frac{2M}{r}+\frac{Q^2}{r^2}-\omega_H^2 r^2 \,,
\end{equation} 
where, as mentioned before,  $\omega_H$ is the de Sitter Hubble constant in our notation.  The corresponding frame with Painlev\' e coordinates and metric (\ref{ss}),  $\{t,r,\theta,\phi\}_{BH}$, has the asymptotic behavior of  the de Sitter comoving  frame $\{t,r,\theta,\phi\}$ with
\begin{equation}
f(r) \to f_0(r)=1-\omega_H^2 r^2\,.
\end{equation}
For this reason we say that  $\{t,r,\theta,\phi\}_{BH}$ is the {\em comoving}  frame of the Reissner-Nordstrom black hole in the de Sitter expanding universe. In the asymptotic zone the frame $\{t,r,\theta,\phi\}_{BH}$ can be related to the observer comoving one, $\{t,r,\theta,\phi\}$,  through de Sitter isometries \cite{CdSR1}.  We assume that the observers stay at rest in the origins of their own  comoving frames evolving along the unique time-like Killing vector field of the de Sitter geometry which  is not time-like everywhere but has this property just in the null cone where the observations are allowed \cite{CGRG}.

The black hole horizons in its proper frame, $\{t,r,\theta,\phi\}$ are given by  the solutions of Eq.  $f(r)=0$ which cannot be written in algebraic closed forms. Therefore,  we may resort to a short numerical analysis for which  it is convenient to introduce the dimensionless coordinate $\rho$ and the new parameters $\kappa$ and $\mu$ defined as 
\begin{equation}\label{par}
\rho=\frac{r}{M}\,,~~~~ q=\frac{Q}{M}\,,~~~~\mu=\omega_H M=\frac{M}{l_H}\,,
\end{equation}
allowing us to write
\begin{equation}\label{fro}
f(r)\to f(\rho)=1-\frac{2}{\rho}+\frac{q^2}{\rho^2}-\mu^2\rho^2\,.
\end{equation}
When $\mu=0$ the genuine Reissner-Nordstrom black hole has two  horizons at 
\begin{equation}
r_{\pm}=M\rho_{\pm}\,, \quad \rho_{\pm}=1\pm\sqrt{1-q^2}\,,
\end{equation}
such that  the condition $0\le |q|\le 1$ becomes mandatory. For $|q|=1$ we obtain the extremal black hole whose horizons are degenerated. 

When we add the de Sitter gravity we must take into account that the parameter $\mu$ is extremely small  such that the black hole horizons remain very close to the values  $r_{\pm}$ while the specific de Sitter horizon is very far, near $r_{dS}\sim \omega_H^{-1} \to \rho_{dS}\sim \mu^{-1}$. Now the black hole with $q=1$ has two distinguishing  horizons at
\begin{eqnarray}
\hat\rho_-&=&\frac{\sqrt{1+4\mu}-1}{2\mu}\,,\\
 \hat\rho_+&=&\frac{1-\sqrt{1-4\mu}}{2\mu}\,,
\end{eqnarray}
but which remain very close each other when $\mu\ll 1$ since  $\hat\rho_+-\hat\rho_-=2\mu+{\cal O}(\mu^2)$. This means that now the range of $q$ is larger but our numerical examples indicate that the upper limit is so close to $1$ such that it is convenient to keep the condition $|q|\le 1$ as long as $\mu$ remains small.  For example, even for larger values as $\mu=0.01$ the extremal black hole is obtained for $q=1.005$ as we can see in Fig.1. Note that in this figure as in the following two we consider exaggerated high values of $\mu$ and $q$ since otherwise the differences between the charged and neutral black holes cannot be pointed out graphically. Similar values are used currently in numerical simulations from the same reasons \cite{sim}. 

\section{Spiral geodesics}

The shape of  the null geodesics in the equatorial plane (with $\theta=\frac{\pi}{2}$ ) of the black hole frame $\{t,r,\theta,\phi\}_{BH}$ are given by  the functions $r(\phi)$ which satisfy the equation \cite{CBH}
\begin{equation}\label{Bin1}
\left(\frac{dr(\phi)}{d\phi}\right)^2-r(\phi)^4\frac{E_{ph}^2}{L_{ph}^2}+r(\phi)^2f[r(\phi)] 
=0\,,
\end{equation}
resulted from the conservation of the photon energy, $E_{ph}$, and angular momentum along the third axis, $L_{ph}$.  Note that this equation is the same as in the static frame since this is static, giving only the shape of trajectory in the same space coordinates.  In fact, the time evolution on geodesics is quite different in the static and comoving frames \cite{Gib}.   

In what follows it is convenient to consider the function $\rho(\phi)$ and the new parameters defined by Eq.  (\ref{par}) for bringing  Eq. (\ref{Bin1}) in homogeneous form
\begin{equation}\label{Bin2}
\left(\frac{d\rho(\phi)}{d\phi}\right)^2-\rho(\phi)^4\frac{E_{ph}^2 M^2}{L_{ph}^2}+\rho(\phi)^2  f[\rho(\phi)] 
=0\,,
\end{equation}
where $f(\rho)$ is given by Eq. (\ref{fro}).
This equation  has two types of solutions, namely circular geodesics on the photon sphere and the associated spiral ones. 

We start with the circular geodesics which must satisfy simultaneously the conditions 
\begin{equation}
\frac{d\rho(\phi)}{d\phi}=\frac{d^2\rho(\phi)}{d\phi^2}=0\,.
\end{equation}
giving a system of two algebraic equations whose solutions are the radius of the photon sphere
\begin{equation}
r_{ph}=\kappa M\,, \quad \kappa =\frac{3}{2}+\frac{1}{2}\sqrt{9-8\,q^2}\,, 
\end{equation}
derived in Ref. \cite{RN} and the mandatory condition 
\begin{equation}\label{Lps}
L_{ph}=\pm\frac{\kappa M E_{ph}}{\sqrt{\lambda^2 - \kappa^2\mu^2}}\,, \quad \lambda^2=1-\frac{2}{\kappa}+\frac{q^2}{\kappa^2}\,.
\end{equation}
The new parameters introduced above have restricted ranges
\begin{equation}
2\le  \kappa\le 3\,, \qquad \frac{1}{4}\le \lambda^2 \le \frac{1}{3}\,,
\end{equation}
since $q\in [0,1]$.

Furthermore, we substitute the condition (\ref{Lps}) in Eq. (\ref{Bin2}) obtaining the new equation
\begin{eqnarray}
&&\left(\frac{d\rho(\phi)}{d\phi}\right)^2=\frac{(\rho(\phi)-\kappa)^2}{2\kappa^3}\nonumber\\
&&~~~~~~\times\left[(\kappa-1)(\rho(\phi)+\kappa)^2-2\kappa^2\right]\,,~~~~\label{eqph}
\end{eqnarray} 
which is independent on the Hubble de Sitter constant $\omega_H$ as in the case of the Schwarzschild-de Sitter black holes \cite{CBH}.  This equation has three constant solutions giving circular geodesics. Apart from the double solution $\rho_{ph}=\kappa$ of the circular geodesics of the photon sphere there are more two constant solutions 
\begin{eqnarray}
\rho_1&=&-\kappa\frac{\sqrt{\kappa-1}-\sqrt{2}}{\sqrt{\kappa-1}}\,,\\
\rho_2&=&-\kappa\frac{\sqrt{\kappa-1}+\sqrt{2}}{\sqrt{\kappa-1}}\,,
\end{eqnarray}
but which do not make sense since $\rho_1<1$ falls inside the exterior horizon and $\rho_2$ is negative.
More surprising is that Eq. (\ref{eqph}) can be integrated analytically obtaining  the shapes of the spiral geodesics $r(\phi)=M\rho(\phi)$ given by the function  
\begin{eqnarray}
&&\rho(\phi)= \frac{\kappa}{\cosh \nu(\phi-\phi_0)-\sqrt{2\kappa-2}\,}\nonumber\\
&&\times\left[\cosh \nu(\phi-\phi_0)+\sqrt{2\kappa-2}\,\frac{\kappa-2}{\kappa-1} \right]
\,,~~~~\label{spir}
\end{eqnarray} 
depending on the arbitrary integration constant  $\phi_0$ and the parameter
\begin{equation}\label{nu}
\frac{1}{\sqrt{2}}\le \nu=\sqrt{2-\frac{3}{\kappa}}\le 1\,.
\end{equation}
Obviously, these functions are determined up to a rotation  fixing the origin of the angular coordinate. 

{ \begin{figure}
 \centering
   \includegraphics[scale=0.95]{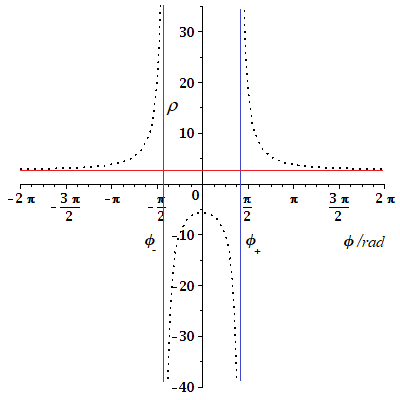}
 \caption{The function $\rho(\phi)$ defined by Eq. (\ref{spir}) with $q=0.27$, $\kappa=2,95$, $\nu=0.991$, $\phi_0=0$ and  vertical asymptotes at $\phi_{\pm}=\pm 1.313\, rad$.}
  \end{figure}}

As the functions (\ref{spir}) are derived here for the first time it deserves to inspect briefly their properties in the simpler case of $\phi_0=0$. Then the function is symmetric, $\rho(\phi)=\rho(-\phi)$, having a pair of symmetric vertical asymptotes at 
\begin{equation}\label{ang}
\phi_{\pm}=\pm\sqrt{\frac{\kappa}{2\kappa-3}}\,{\rm arccosh} \sqrt{2\kappa-2}\,.
\end{equation}
Moreover, the limits 
\begin{equation}
\lim_{\phi\to\pm \infty}\rho(\phi)=\kappa\,,
\end{equation}
point our the horizontal asymptote corresponding to the radius of the photon sphere.
In Fig. 2 we see that  this function makes sense only on the domain $(-\infty, \phi_-)\cup (\phi_+,\infty)$ where this remains outside the photon sphere,  $\rho(\phi)\ge \kappa$. The domain $[\phi_-,\phi_+]$ is an opaque window where the function   is negative having no physical meaning.   

It is remarkable that all the results concerning the spiral geodesics are independent on $\mu$ which is the only parameter depending on the de Sitter gravity. This means that we find similar spiral geodesics around the Reissner-Nordstrom black hole as in Minkowski spacetime where $\omega_H=0 ~\to~\mu=0$. In fact the de Sitter gravity is encapsulated only in Eq. (\ref{Lps}) giving the crucial quantity in determining the redshift and the black hole shadow.

Finally, we observe that for  $q= 0$ we have $\kappa=3$,  $\nu=1$ and $\lambda^2=\frac{1}{3}$ recovering thus the results obtained for the Schwarzschild-de Sitter black hole \cite{CBH}, namely the radius of the photon sphere $\hat r_{ph}=3M$ and the condition 
\begin{equation}\label{Lps1}
\hat L_{ph}=\pm\frac{3\sqrt{3} M E_{ph}}{\sqrt{1 - 27 \mu^2}}\,, 
\end{equation}
derived in Ref. \cite{SS1}, while the functions (\ref{spir})  become 
\begin{equation}
\hat\rho(\phi)=3\,\frac{\cosh( \phi-\phi_0) +1}{\cosh(\phi-\phi_0) -2}\,.
\end{equation}
Note that this function is the same as in the case of the Schwarzschild black holes in Minkowski spacetime, complying with Darwin's formula \cite{D1},
\begin{eqnarray}
\frac{1}{r(\phi)}&=&\frac{1}{\hat\rho(\phi)M}\nonumber\\
&=&-\frac{1}{6M}+\frac{1}{2M}\, \tanh^2 \left(\frac{\phi-\phi_0}{2}\right)\,,~~~~~
\end{eqnarray}
since, as in the general case, the parameter $\omega_H$ arises only in Eq. (\ref{Lps1}) which was used for deriving the results of Refs. \cite{CBH,Comp}.         

{ \begin{figure}
 \centering
   \includegraphics[scale=0.6]{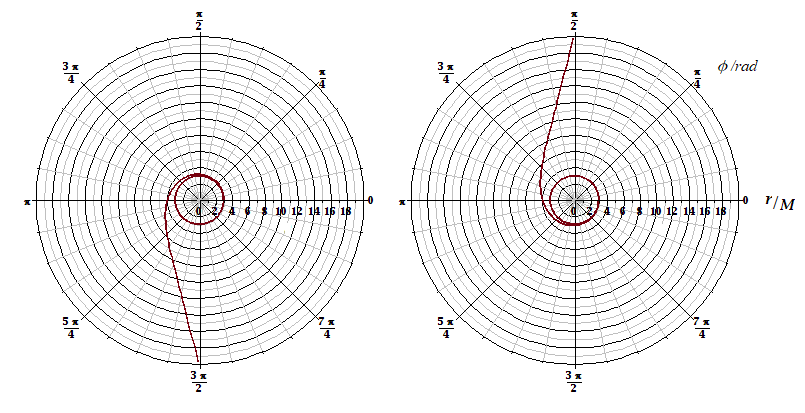}
 \caption{The spiral geodesics given by the function (\ref{spir}) with the same parameters as in Fig. 2,  for $\phi\in[-4\pi,\phi_-]$ (left panel) and $\phi\in[\phi_+, 4\pi]$ (right panel). }
  \end{figure}}

\section{Shadow and redshift}

The spiral geodesics are the closest trajectories to the photon sphere of the first photons that can be observed at the limit of the black hole shadow. Therefore, for studying this shadow and the associated  redshift we have to consider only these photons that, in general, may be emitted by a moving  black hole and measured by a fixed observer. 

The photon is emitted in the black hole comoving frame with  the  momentum ${\bf k}$, energy $E_{ph}=k=|{\bf k}|$ and angular momentum  (\ref{Lps}). This is observed as coming from an apparent source $S$ situated on a sphere of radius
\begin{equation}\label{rS}
r_S=\frac{|L_{ph}|}{k}=\rho_SM\,,
\end{equation} 
depending on the new parameter  
\begin{eqnarray}
\rho_S&=&\frac{\kappa} {\sqrt{\lambda^2-\kappa^2\mu^2}}\nonumber\\
&=& \frac{\kappa}{\lambda}+\frac{1}{3}\frac{\kappa^3}{\lambda^3}\,\mu^2+{\cal O}(\mu^4)\,.\label{fS}
\end{eqnarray} 
The apparent trajectory of this photon is a de Sitter geodesic whose conserved quantities depend exclusively on ${\bf k}$ and $r_S$ \cite{CBH}.  Then if we know  the relative motion of the black hole with respect to the fixed observer we may derive  the conserved quantities measured by this observer  by using suitable isometries of the de Sitter relativity \cite{CdSR1,CDop}. 

In Ref. \cite{CBH} we assumed that the photon is emitted at the initial moment $t=0$  when the black hole is translated with  the distance $d=\delta M$ moving along this direction with the peculiar velocity $V$  with respect to the fixed observer.  Then by using a translation followed by a Lorentzian isometry   we deduced the general formulas of the observed quantities from which we extracted the redshift and shadow of the Schwarzschild-de Sitter black holes in terms of $d$, $V$ and the specific parameter  $\hat\xi=\frac{\hat r_S}{d}$ which depends on the black hole geometry only trough $\hat r_S=\hat\rho_S M=\frac{|\hat L_{ph}|}{k}$ where $\hat L_{ph}$ is given by Eq. (\ref{Lps1}) \cite{CBH}.

For deriving the similar results in the case of our Reissner-Nordstrom-de Sitter black hole we have to take over all the results presented in Refs. \cite{CBH,Comp} substituting the parameter $\hat\xi$ with the new one, 
\begin{equation}\label{xirs}
\xi=\frac{r_S}{d}=\frac{\rho_S}{\delta}\,,
\end{equation} 
given by the actual radius (\ref{fS}).  For example, in the particular case when the black hole does not have an initial relative velocity ($V=0$) we may write the simple formulas \cite{CBH}
\begin{eqnarray}
\sin\alpha&=&\xi\,, \label{fina}\\
\frac{1}{1+z}&=&1-\mu\,\delta \cos\alpha\,,\label{finz}
\end{eqnarray}
showing how the observations of the shadow angular radius $\sin\alpha$ and the redshift $z$ are related each other.

The results presented here cover three particular  cases, namely the Schwarzschild black hole ($q=0$) in Minkowski ($\mu=0$) \cite{Sy} and de Sitter ($\mu\not=0$) \cite{SS1,CBH} spacetimes or the Reissner-Nordstrom black hole ($q\not=0$) in Minkowski spacetime ($\mu=0$).

We have seen that in the simplest case of the Schwarzschild black hole in Minkowski spacetime  we recover the well-known radius of the photon sphere $\kappa =3$  and the parameter $\hat\rho_S=3\sqrt{3}$ giving the black hole shadow \cite{CBH}. In the case of  charged black holes in de Sitter gravity the corresponding parameter (\ref{fS}) is a complicated function of $q=\frac{Q}{M}$ and $\mu=\frac{M}{l_H}$. We observe first that $\mu$  remains very small in our actual expanding universe. For example, even in the case of a very massive black hole of mass $M=10^9 M_{\odot}$, this is of the order  $10^{-14}$.  Therefore, when we look for charged black holes we may neglect the terms of the order ${\cal O}(\mu^2)$ in Eq. (\ref{fS}) using the  approximative formula 
\begin{equation}\label{rnum}
\rho_S\sim 3\sqrt{3}-\frac{\sqrt{3}}{2}\,q^2-\frac{7\sqrt{3}}{27}q^4\,,
\end{equation}
which assures a satisfactory accuracy for $q<0.3$. In fact,  the first correction of the order ${\cal O}(q^2)$ is enough for finding  rapidly charged black holes based on the observed black hole shadow and redshift whether the mass $M$ (and implicitly $\mu$) is known. Indeed, deriving first the distance $\delta$ from  Eq.  (\ref{finz}) and  combining then  Eqs, (\ref{xirs}) and (\ref{fina}) we find the parameter
\begin{equation}\label{rzum}
\rho_S=\frac{\tan\alpha}{\mu}\,\frac{z}{1+z}\,,
\end{equation} 
which has to be less than $3\sqrt{3}$ if the black hole carries electric charge. We obtain thus a criterion of detecting charged black holes but which is delicate depending on the accuracy of the data we use.

\section{Concluding remarks}

We derived in premier the functions giving the shapes of  the spiral geodesics around the Reissner-Nordstrom black holes in the de Sitter expanding universe and the parameter $\xi$ giving the redshift and black hole shadow according to our method we proposed recently \cite{Comp,CBH}. Thus we obtain a framework in which one can detect the charged black holes with $V=0$  by testing whether the astronomical data match with  Eqs. (\ref{rnum}) and (\ref{rzum}). For the black holes having peculiar velocities we may apply the same method manipulating the more complicated formulas  by using our computer code \cite{Comp}. However, in both these cases some doubts concerning the accuracy of the observation data as well as the associated optical phenomena due to the cosmic dust and plasma may lead to some uncertainty in interpreting the final results. 

Technically speaking, the method we apply here  is based on the transformation rules under isometries of the conserved quantities being thus independent on the coordinates we use, static or Painlev\' e ones. Nevertheless, we preferred the comoving frames with  Painlev\' e coordinates for keeping under control the philosophy of  the remote observers and the black hole relative motion in the de Sitter spacetime. Moreover, in this frame $d$ is just the physical distance between observer and black hole at the time when this emits the  light.   

In general, the method based on the study of the conserved quantities is not enough for understanding the entire information carried out by the light emitted by static or moving black holes. There are important quantities that can be deduced from the coordinate transformations under isometries as, for example, the photon propagation time or the real distance between observer and black hole at the time when the photon is measured \cite{CDop}. For this reason we hope that our algebraic method will improve the general geometric approach for getting over the difficulties in analysing the light emitted by various cosmic objects in the de Sitter expanding universe or other geometries.


\begin{thebibliography}{}


\bibitem{Sy} 
J. L. Synge, {\em Mon. Not. R. Astron. Soc.} {\bf 131} (1966) 463.

\bibitem{SS1}
V. Perlick, O. Yu. Tsupko, G. S. Bisnovatyi-Kogan, {\em Phys. Rev. D} {\bf 97} (2018) 104062.

\bibitem{SS2}
G. S. Bisnovatyi-Kogan and O. Yu. Tsupko,  {\em Phys. Rev. D} {\bf 98} (2018) 084020. 

\bibitem{SS01}
J. T. Firouzjaee and A. Allahyari, {\em Eur. Phys. J. C} {\bf 79} (2019) 1140.

\bibitem{SS02}
Z. Chang and Q.-H. Zhu, {\em JCAP} {\bf 06} (2020) 055.

\bibitem{SS03}
S. Vagnozzi, C. Bambi, and L. Visinelli, {\em Class. Quantum Grav.} {\bf 37} (2020)  087001. 

\bibitem{SS3}
O. Yu. Tsupko G. S. Bisnovatyi-Kogan, {\em  Int.  J.  Mod. Phys. D} {\bf 29} (2020) 2050062. 

\bibitem{K1}
J. M. Bardeen, {\em Proceedings, Ecole d'Et de Physique Thorique: Les Astres Occlus.} (Les Houches, France, 1973).

\bibitem{K2}
H. Falcke, F. Melia, and E. Agol, {\em Astrophys. J. Lett.} {\bf 528} (1999) L13.

\bibitem{KdS1}
A. Grenzebach, V. Perlick, and C. L\" ammerzahl, {\em Phys. Rev. D} {\bf 89} (2014) 124004.
\bibitem{KdS2}
Z. Stuchlik, D. Charbul\' ak, and J. Schee, {\em Eur. Phys. J. C} {\bf 78} (2018) 180.

\bibitem{KdS3}
P.-C. Li, M. Guo, and B. Chen, {\em Phys. Rev. D} {\bf 101} (2020) 084041. 

\bibitem{KdS4}
Z. Chang and Q.-H. Zhu, {\em Phys. Rev. D} {\bf 101} (2020) 084029.

\bibitem{V1}
C. Bambi, K.  Freese,  S. Vagnozzi and L. Visinelli, {\em Phys. Rev. D} 100 (2019) 044057.

\bibitem{V2}
S. Vagnozzi and L. Visinelli, {\em Phys. Rev. D} {\bf 100} (2019) 024020.

\bibitem{A1}
K. Akiyama, et al., {\em Astrophys. J.} {\bf 875}(1) (2019) L1.  
\bibitem{A2}
K. Akiyama, et al., {\em Astrophys. J.} {\bf 875}(1) (2019) L6. 

\bibitem{L1}
G. E. Lema\^ itre, {\em Ann. Soc. Sci. de Bruxelles} {\bf 47A} (1927) 49.

\bibitem{L2}
G. E. Lema\^ itre, {\em MNRAS} {\bf 91} (1931) 483.

\bibitem{Hubb}
E. Hubble, {\em Proc. Nat. Acad. Sci.} {\bf 15} (1929) 168.

\bibitem{SW}
S. Weinberg, {\em Gravitation and Cosmology: Principles and Applications of the General Theory of relativity} (J. Wiley \& Sons, New York 1972). 

\bibitem{H0}
E. R. Harrison, {\em  Cosmology: The Science of the Universe} (New York: Cambridge Univ. Press, 1981).

\bibitem{H}
E. Harrison, {\em  Astrophys. J.} {\bf 403} (1993) 28.

\bibitem{LL}
L. D. Landau and E. M. Lifshitz, {\em The classical theory of fields} (Elsevier Sci. Inc. NY. 1975).

\bibitem{CdSR1}
I. I. Cot\u aescu, {\em Eur. Phys. J. C} {\bf 77} (2017) 485.

\bibitem{CdSR2}
I. I. Cot\u aescu, {\em Eur. Phys. J. C} {\bf 78}  (2018) 95.

\bibitem{CDop}
I. I. Cot\u aescu, {\em Mod. Phys. Lett. A} {\bf  36} (2021) 2150022.

\bibitem{CBH}
I. I. Cot\u aescu, {\em Eur. Phys. J. C.} {\bf 81} (2021) 32.  

\bibitem{Comp}
I. I. Cot\u aescu,  {\em Maple code BH01} (2020) download here \href{https://physics.uvt.ro/~cota/CCFT/codes}{BH01}

\bibitem{D1}
 C. Darwin, Proc. Roy. Soc. {\bf 249} (1958) 180.
 
\bibitem{D2}
 C. Darwin, Proc. Roy. Soc. {\bf 263} (1961) 39.

\bibitem{RN}
M. Mokdad, {\em Class. Quantum Grav.} {\bf 34} (2017) 175014. 

\bibitem{Pan}
P. Painlev\' e, {\em  C. R. Acad. Sci.} (Paris) {\bf 173} (1921) 677.

\bibitem{BD}
N. D. Birrel and P. C.W. Davies,  {\em Quantum Fields in Curved Space} (Cambridge University Press, Cambridge 1982).

\bibitem{CGRG}
I. I. Cot\u aescu, {\em GRG} {\bf 43} (2011) 1639.

\bibitem{sim}
G. Bozzola  and V. Paschalidis, {\em Phys. Rev. D} {\bf 99} (2019) 104044.

\bibitem{Gib}
G. W. Gibbons, C. M. Warnick, and M. C. Werner, {\em Class. Quantum Grav.} {\bf 25} (2008) 245009.







\end{thebibliography}
\end{document}